\documentclass{PoS}

\usepackage{bm}

\newcommand{\ahvp}{a_\mu^{\rm hvp}}
\newcommand{\ahlbl}{a_\mu^{\rm hlbl}}
\newcommand{\Nf}{N_{\rm f}}

\newcommand{\be}{\begin{equation}}
\newcommand{\ee}{\end{equation}}
\newcommand{\bea}{\begin{eqnarray}}
\newcommand{\eea}{\end{eqnarray}}

\title{The hadronic vacuum polarization contribution to $(g-2)_\mu$
  from 2\,+\,1 flavours of O(\bm{$a$}) improved Wilson quarks}

\ShortTitle{HVP contribution to $(g-2)_\mu$ with Wilson quarks}

\author{Antoine G\'erardin$^{1,2,3}$,
  Tim Harris$^{2,4}$, Georg von Hippel$^{1}$, Ben H\"orz$^{1}$,
  Harvey Meyer$^{1,2}$, Daniel Mohler$^{1,2}$, Konstantin Ottnad$^{1}$,
  \speaker{Hartmut Wittig}$^{\,1,2}$ \\ 
        \llap{$^1$}PRISMA Cluster of Excellence and Institute for Nuclear
        Physics, University of Mainz, D-55099 Mainz, Germany\\
        \llap{$^2$}Helmholtz Institute Mainz, University of Mainz, D-55099 Mainz,
        Germany\\
        \llap{$^3$}John von Neumann Institute for Computing, DESY,
        Platanenallee 6, D-15738 Zeuthen, Germany\\ 
        \llap{$^4$}Dip. di Fisica G. Occhialini, Universit\`a di Milano-Bicocca,
        and INFN, Sezione di Milano-Bicocca, Piazza della Scienza 3,
        I-20126 Milano, Italy \\
        E-mail:
        \email{antoine.gerardin@desy.de}, \email{tim.harris@mib.infn.it}, \email{\{hippel,hoerz,meyerh,damohler,kottnad,wittigh\}@uni-mainz.de}}

\abstract{We report on our ongoing project to determine the
  leading-order hadronic vacuum polarisation contribution to the muon
  $g-2$, using ensembles with $\Nf=2+1$ flavours of O($a$) improved
  Wilson quarks generated by the CLS effort, with pion masses down to
  the physical value. We employ O($a$) improved versions of the local
  and conserved vector currents to compute the contributions of the
  light, strange and charm quarks to $(g-2)_\mu$, using the
  time-momentum representation. We perform a detailed investigation of
  the systematic effects arising from constraining the long-distance
  regime of the vector correlator. To this end we make use of
  auxiliary calculations in the iso-vector channel using distillation
  and the L\"uscher formalism. Our results are corrected for
  finite-volume effects by computing the timelike pion form factor in
  finite and infinite volume. For certain parameter choices, the
  corrections computed in this way can also be confronted with results
  determined on different volumes. Currently, the overall precision of
  our results is limited by the uncertainties in the lattice scale.}

\FullConference{The 36th Annual International Symposium on Lattice Field Theory - LATTICE2018\\
		22-28 July, 2018\\
		Michigan State University, East Lansing, Michigan, USA.}

\begin{document}

\section{Introduction}

The persistent discrepancy of $\approx3.5$ standard deviations between
the direct measurement of the muon's anomalous magnetic moment
$a_\mu=\frac{1}{2}(g-2)_\mu$ and the value predicted by theory
\cite{Tanabashi:2018oca} is one of the most promising hints for
physics beyond the Standard Model. In view of the fact that the
experimental precision is set to increase by a factor four, owing to
the efforts by the Fermilab E989 and J-PARC E34 experiments, it is of
paramount importance to reduce the error of the theoretical
prediction. Lattice QCD calculations have emerged as a promising
method to obtain estimates for both the hadronic vacuum polarisation
and light-by-light scattering contributions, $\ahvp$ and $\ahlbl$,
with controlled uncertainties (see the recent review in
\cite{Meyer:2018til}). In Ref.~\cite{DellaMorte:2017dyu} we published
a result for $\ahvp$ in two-flavour QCD, i.e.
\be
   \ahvp = (654\pm32\,{^{+21}_{-23}})\cdot10^{-10},
\ee
where the first error is statistical and the second represents a
combination of various systematic errors. Clearly, the total precision
of $\approx6$\,\% is insufficient to have an impact on resolving the
$g-2$ puzzle. Here we report on our ongoing effort to determine
$\ahvp$ in lattice QCD with $\Nf=2+1$ flavours of O($a$) improved
Wilson fermions. Previous accounts have been given in
\cite{DellaMorte:2017khn} and \cite{Meyer:2018til}.

\section{Features of our calculation}

Our evaluation of $\ahvp$ is based on the time-momentum representation
(TMR) \cite{Bernecker:2011gh}, i.e.
\be\label{eq:TMR}
  \ahvp = \left(\frac{\alpha}{\pi}\right)^2
  \int_0^{\infty}dx_0\,\tilde{K}(x_0) G(x_0),
\ee
where $\tilde{K}(x_0)$ is a known kernel function depending on the
Euclidean time $x_0$ \cite{Bernecker:2011gh,DellaMorte:2017dyu}, and
$G(x_0)$ is the spatially summed correlator of two electromagnetic
currents
\be
  G(x_0)=-a^3\sum_{\vec{x}}\,\left\langle J_k(x)J_k(0) \right\rangle,
  \quad
  J_\mu(x) = {\textstyle\frac{2}{3}}\bar{u}\gamma_\mu u
            -{\textstyle\frac{1}{3}}\bar{d}\gamma_\mu d
            -{\textstyle\frac{1}{3}}\bar{s}\gamma_\mu s + \ldots.
\vspace{-0.2cm}
\ee
Previous calculations suggest that the tail of the integrand
$\tilde{K}(x_0)G(x_0)$, i.e. the region $x_0\gtrsim3$\,fm makes a
sizeable contribution of about 3\% to $\ahvp$. Hence, in order to
determine $\ahvp$ with sufficient precision, it is important to
accurately constrain the long-distance regime of the correlator
$G(x_0)$, which is dominated by the two-pion contribution to the
iso-vector component.

Below we sketch the main features of our calculation. We employ the
gauge ensembles generated as part of the CLS effort, using $\Nf=2+1$
flavours of O($a$) improved Wilson quarks \cite{Bruno:2014jqa}. In
order to combat the problem of topology freezing, a large fraction of
the ensembles have open boundary conditions in the time direction
\cite{Schaefer:2010hu,Luscher:2011kk}. The hopping parameters of the
dynamical light and strange quarks were chosen such that the physical
point is reached along a trajectory where ${\rm Tr}\,M = \rm
const$. Further details of the 17 ensembles used in this study are
listed in Table\,\ref{tab:ensembles}: we have computed $\ahvp$ for
four different lattice spacings, covering a range of pion masses from
around 400\,MeV down to the physical one. While all ensembles satisfy
$m_\pi L\geq4$, we can also study finite-volume effects by comparing
results at different volumes for fixed $m_\pi$. The lattice scale is
set via the gradient flow time, which has been determined in
\cite{Bruno:2016plf} as $\sqrt{8t_0}=0.415(4)(2)$\,fm. The current
uncertainty of the scale determination of about 1\% limits the overall
precision of our calculation of $\ahvp$ (see the discussion on the
scale setting error in \cite{DellaMorte:2017dyu,DellaMorte:2017khn}).

In order to constrain the long-distance regime of the vector
correlator, we have performed a dedicated calculation of the
light-quark iso-vector correlator, employing correlator matrices and
the distillation technique \cite{Peardon:2009gh,Morningstar:2011ka},
which allows not only for the determination of the $I=1$ scattering
phase shift but also of the timelike pion form factor
\cite{Meyer:2011um}. For the evaluation of quark-disconnected diagrams
we have resorted to using hierarchical probing
\cite{Stathopoulos:2013aci}, as well as a new covariant coordinate
space technique \cite{Meyer:2017hjv}.

\begin{table}
{\small 
\begin{center}
\begin{tabular}{c c c c c c c c c}
\hline\hline
ensemble & BCs & $\beta$ & $a$\,[fm] & $L/a$ & $T/a$ & $L$\,[fm] &
$m_\pi$\,[MeV] & $m_\pi L$ \\
\hline
H101 & obc & 3.40 & 0.0865  & 32 &  96 & 2.77 & 420 & 5.9 \\
H102 & obc &      &         & 32 &  96 & 2.77 & 350 & 4.9 \\
H105 & obc &      &         & 32 &  96 & 2.77 & 280 & 3.9 \\
N101 & obc &      &         & 48 & 128 & 4.15 & 280 & 5.9 \\
C101 & obc &      &         & 48 &  96 & 4.15 & 220 & 4.6 \\
\hline
B450 & pbc & 3.46 & 0.0765  & 32 &  64 & 2.45 & 415 & 5.2 \\
S400 & obc &      &         & 32 & 128 & 2.45 & 350 & 4.3 \\
N401 & obc &      &         & 48 & 128 & 3.67 & 280 & 5.2 \\
\hline
H200 & obc & 3.55 & 0.0644  & 32 &  96 & 2.06 & 420 & 4.4 \\
N202 & obc &      &         & 48 & 128 & 3.09 & 420 & 6.6 \\
N203 & obc &      &         & 48 & 128 & 3.09 & 340 & 5.3 \\
N200 & obc &      &         & 48 & 128 & 3.09 & 280 & 4.4 \\
D200 & obc &      &         & 64 & 128 & 4.12 & 200 & 4.2 \\
E250 & pbc &      &         & 96 & 192 & 6.18 & 135 & 4.2 \\
\hline
N300 & obc & 3.70 & 0.0499  & 48 & 128 & 2.40 & 420 & 5.1 \\
N302 & obc &      &         & 48 & 128 & 2.40 & 340 & 4.1 \\
J303 & obc &      &         & 64 & 192 & 3.19 & 260 & 4.2 \\
\hline\hline
\end{tabular}
\caption{Simulation parameters and pion masses of the ensembles used
  in this study. The labels ``obc'' and ``pbc'' indicate the use of
  open or periodic boundary conditions,
  respectively. \label{tab:ensembles}}
\end{center}
}
\end{table}

In our calculations we employ the O($a$) improved versions of the
local (loc) and conserved (con) vector currents. For the local
current, $V_\mu^{\rm loc}(x)=(\bar\psi\gamma_\mu\psi)(x)$, the
expression is
\be
  (V_\mu^{\rm loc}(x))^I = V_\mu^{\rm loc}(x)+ 
  ac_{\rm V}^{\rm loc}\,\tilde\nabla_\nu T_{\mu\nu}(x),
\vspace{-0.1cm}
\ee
where $T_{\mu\nu}=-\frac{1}{2}\bar\psi[\gamma_\mu,\gamma_\nu]\psi$ is
the tensor current, $\tilde\nabla_0$ denotes the symmetrised lattice
derivative, and $c_{\rm V}^{\rm loc}$ is an improvement coefficient
that can be evaluated non-perturbatively by imposing chiral Ward
identities \cite{Guagnelli:1997db,Gerardin:2018kpy}. There is an
analogous expression for the improved conserved current, with
coefficient $c_{\rm V}^{\rm con}$. The local vector current must be
renormalised, and in this contribution we employ the non-perturbative
determination of the renormalisation factor $Z_{\rm V}$ from ratios of
three- and two-point correlation functions, similarly to our
two-flavour calculation \cite{DellaMorte:2017dyu}. The expression of
the (mixed) correlator involving the conserved and local currents for
quark flavour~$f$ reads
\be\label{eq:Gfdef}
   G^{lc}_f(x_0) =
   -\frac{a^3}{3}\sum_{k=1}^3\sum_{\vec{x}}{q}_f^2\,Z_{\rm V}
   \left\langle V_{k,f}^{\rm con}(x_0,\vec{x}) V_{k,f}^{\rm loc}(0)
   \right\rangle, \qquad f=ud, s, c, 
\vspace{-0.2cm}
\ee
where $V_{k,f}$ denotes the spatial component of the vector current of
flavour~$f$, and ${q}_f$ is the corresponding electric
charge.\footnote{In the isospin limit we take ${q}_{ud}^2=5/9$.}  Note
that only quark-connected contributions are included in
Eq.\,(\ref{eq:Gfdef}). We also consider the correlator $G^{ll}_f(x_0)$
of two local currents, which contains a factor of $Z_{\rm V}^2$.

A recent analysis of the renormalisation pattern of the improved local
vector current has shown that SU(3) flavour breaking induces a mixing
between the local vector currents of different quark flavours at
O($a$) \cite{Gerardin:2018kpy}. Hence, in order to be consistent with
improvement, the contributions of individual quark flavours to $\ahvp$
must be derived from linear combinations of correlators of
appropriately renormalised quark currents. The findings of
Ref.\,\cite{Gerardin:2018kpy} were not available at the time of the
conference, and we leave the full analysis to a later publication.

\section{Preliminary results}

\begin{figure}[t]
\vspace{-0.4cm}
\centering
\leavevmode
\includegraphics[width=0.49\textwidth,clip]{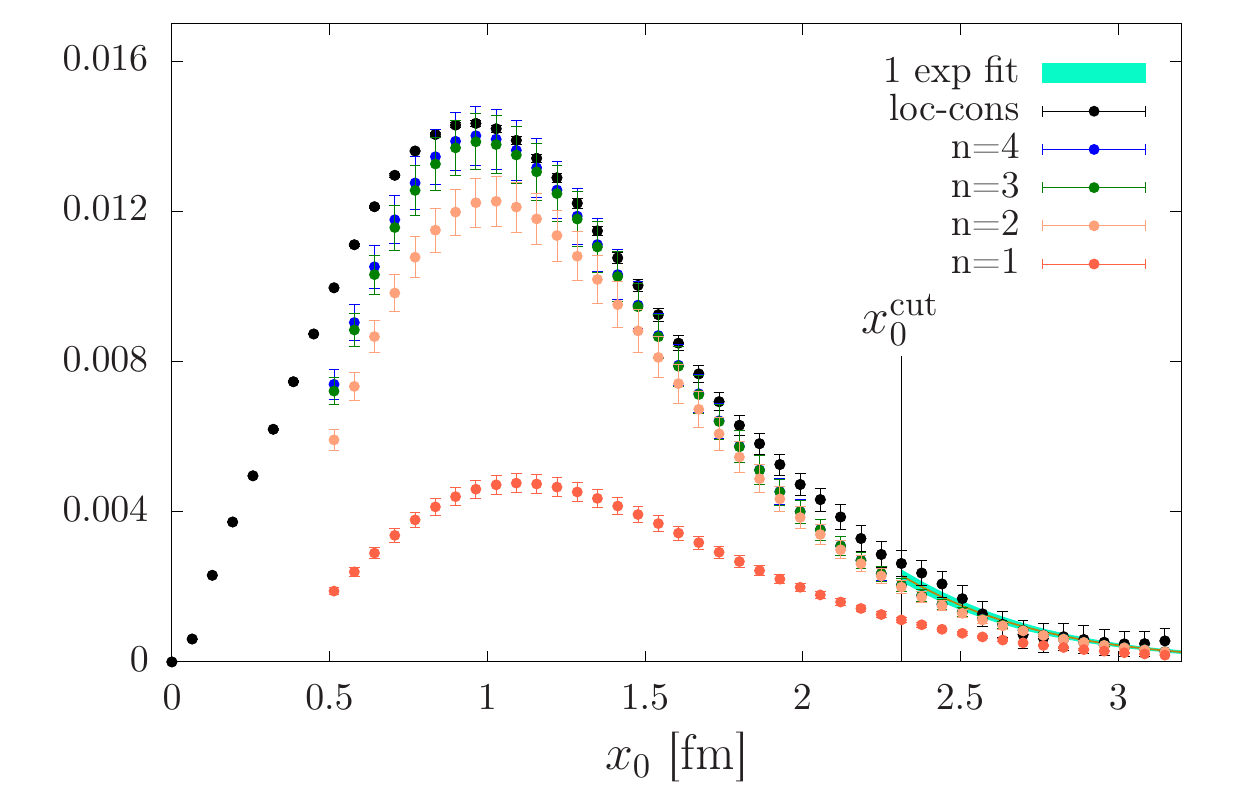}
\hfill
\includegraphics[width=0.49\textwidth,clip]{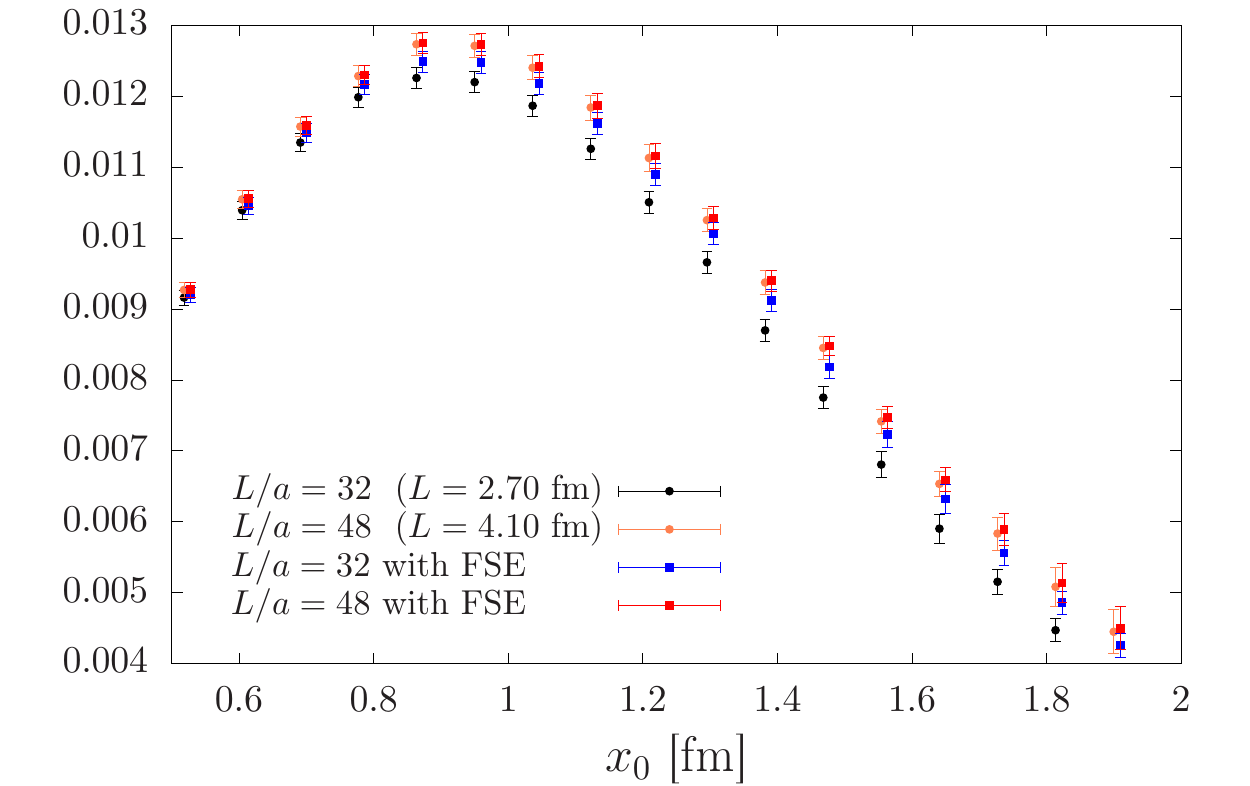}
\vspace{-0.2cm}
\caption{Left: The integrand $\tilde{K}(x_0)G^{ll}_{ud}(x_0)$ on
  ensemble D200 in units of $m_\mu$, compared to the iso-vector
  contribution $\tilde{K}(x_0)G^{\rho\rho}(x_0,L)$. Right: the
  integrand $\tilde{K}(x_0)G^{lc}_{ud}(x_0)$ computed on ensembles
  H105 and N101, before and after applying the finite-volume
  correction computed using the Gounaris-Sakurai parameterisation of
  $F_\pi(\omega)$. \label{fig:integrand}}
\vspace{0.4cm}
\includegraphics[width=0.485\textwidth,clip]{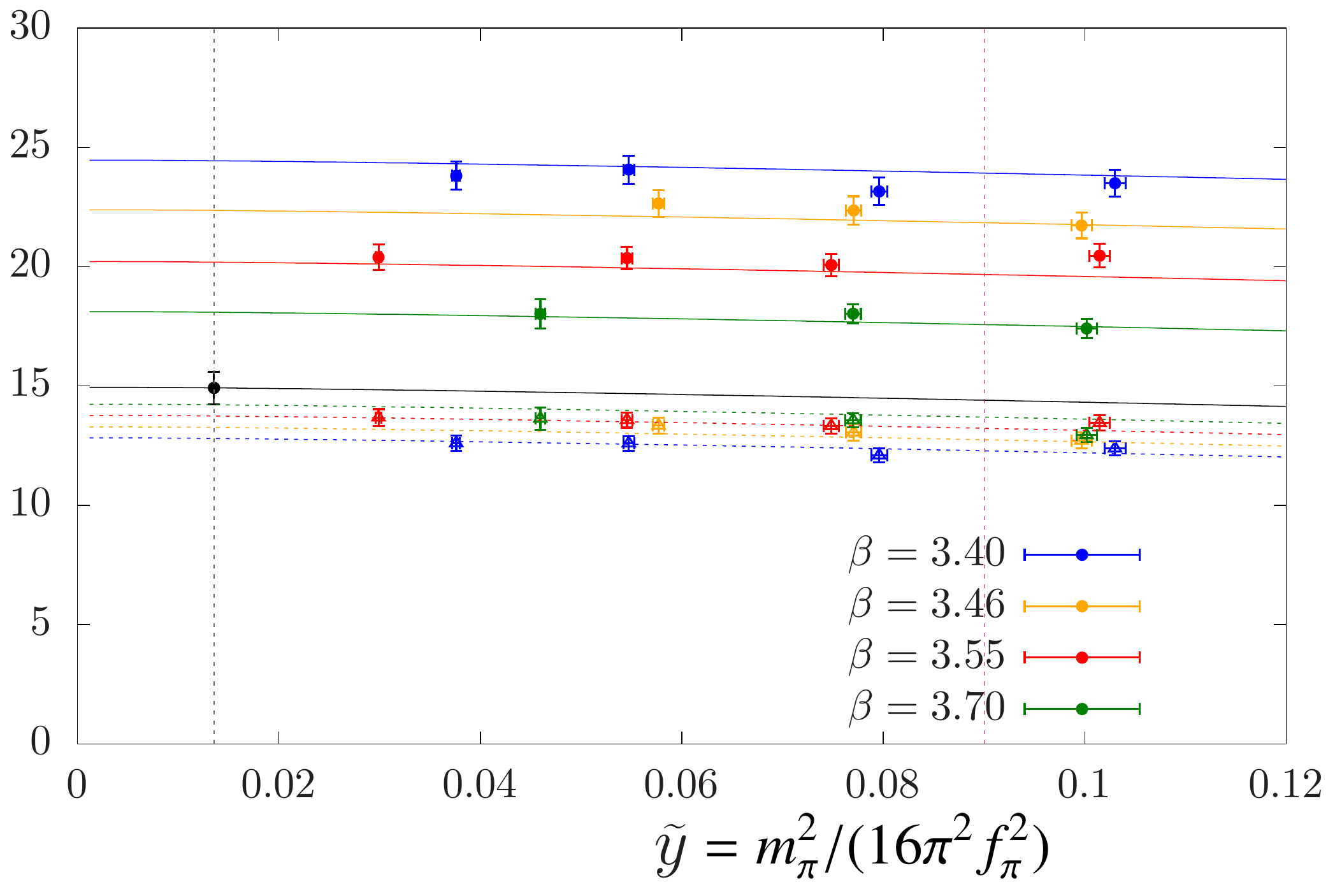}
\hfill
\includegraphics[width=0.49\textwidth,clip]{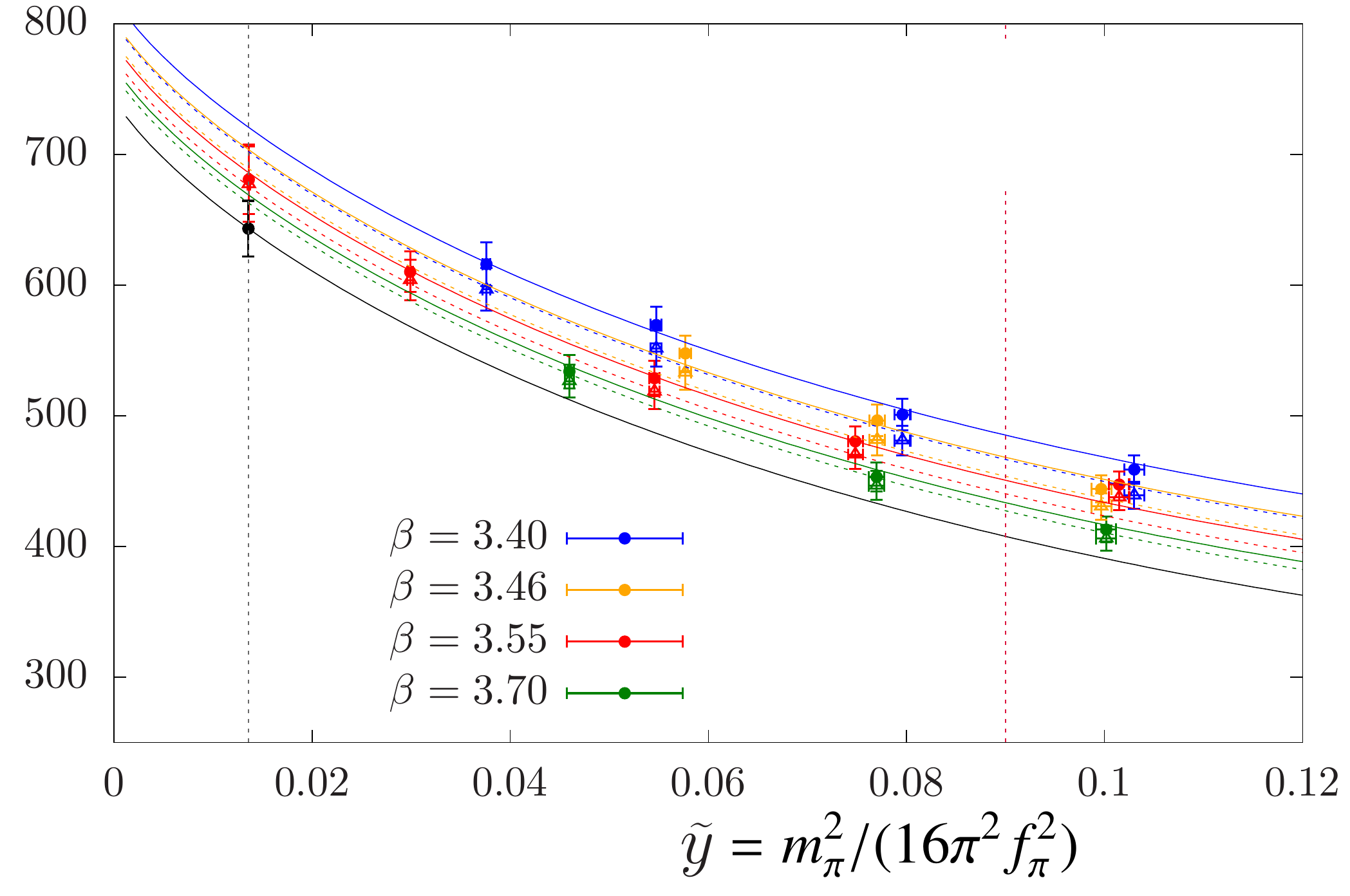}
\vspace{-0.3cm}
\caption{Chiral and continuum extrapolations of the contributions of
  the charm (left) and light quarks (right) to $\ahvp$, in units of
  $10^{-10}$. Data points are plotted versus the variable
  $\tilde{y}=m_\pi^2/(16\pi^2f_\pi^2)$. Solid and dotted lines
  describe the pion mass dependence of the results determined from the
  local-local and local-conserved correlators, respectively. Estimates
  at the physical point are represented by black
  circles.\label{fig:chiralcont}}
\end{figure}

In the left panel of Fig. \ref{fig:integrand} we show the integrand
$\tilde{K}(x_0)G^{ll}_{ud}(x_0)$ on ensemble D200, at a pion mass of
about 200\,MeV. The data, denoted by black points, are compared to the
results of our auxiliary calculation of the iso-vector correlator
$G^{\rho\rho}(x_0,L)$, whose long-distance behaviour is
\be
  G^{\rho\rho}(x_0,L) \stackrel{x_0\gg0}{\sim} \sum_n\,|A_n|^2\,
  {\rm e}^{-\omega_nx_0},\qquad \omega_n=2\sqrt{m_\pi^2+k^2},
\ee
where $k$ is the scattering momentum. The energy levels $\omega_n$
were determined for the four lowest-lying states by computing
correlator matrices using $\rho$ and $\pi\pi$ interpolating operators
and stochastic distillation \cite{Morningstar:2011ka} and solving the
associated generalised eigenvalue problem. In a second step, the
amplitude $A_n\equiv\langle0|V_{k,ud}^{\rm con}(0)|n\rangle$ was
determined as the matrix element of the local current and the
approximate interpolator onto the $n^{\rm{th}}$ energy eigenstate
$|n\rangle$.
From the figure it is clear that the accumulated contributions from
the four lowest-lying states saturate the iso-vector correlator for
$x_0\gtrsim1$\,fm. Furthermore, the long-distance behaviour of the
integrand $\tilde{K}(x_0)G^{ll}_{ud}(x_0)$ is well described by the
iso-vector contribution which is also statistically more
accurate. Finally, one concludes that the two-pion contribution,
denoted by the red filled circles, dominates the vector correlator for
$x_0\gtrsim3$\,fm.

The long-distance behaviour of the vector correlator is closely
related to the important issue of finite-volume effects. In
\cite{Francis:2013qna,DellaMorte:2017dyu} it was shown how
finite-volume corrections can be computed by inserting the difference
of the isovector correlator in finite and infinite volume into the
integral representation Eq.\,(\ref{eq:TMR}). For the latter, the
expression is
\be
  G^{\rho\rho}(x_0,\infty) =
  \int_0^{\infty}d\omega\,\omega^2\,\rho(\omega^2)\,
  {\rm e}^{-\omega|x_0|},\quad\rho(\omega^2)=\frac{1}{48\pi^2} 
  \left(1-\frac{4m_\pi^2}{\omega^2}\right)^{3/2}
  \left|F_\pi(\omega)\right|^2,
\ee
where $\rho(\omega^2)$ denotes the spectral function, which is related
to the timelike pion form factor $F_\pi(\omega)$. In the absence of
any direct lattice calculation of $F_\pi(\omega)$ one may resort to
the Gounaris-Sakurai (GS)
parameterisation\,\cite{Gounaris:1968mw}. Obviously, it is important
to verify the predictions of the GS model by confronting them with
lattice data obtained on different volumes. In the right panel of
Fig.\,\ref{fig:integrand} we compare the TMR integrand computed on
ensembles H105 and N101, which realise the same pion mass
($m_\pi=280$\,MeV) on two different volumes, corresponding to
${m_\pi}L=3.9$ and 5.9, respectively. One then finds that the
finite-volume correction determined by the GS model is tiny for the
larger volume (N101). Second, one finds that the FV-corrected data on
the smaller volume (H105) agree with those on N101 within errors. We
conclude that finite-size effects are well described by the GS
parameterisation of $F_\pi(\omega)$. At the physical pion mass and for
$m_\pi L=4$, which corresponds to our ensemble E250, we estimate a
finite-volume correction to $\ahvp$ of $(20\pm4)\cdot 10^{-10}$ using
the GS model. It is interesting to note that this correction is
reduced by a factor~10 for $m_\pi L=6$. A calculation of the timelike
pion form factor has been performed for a subset of our previously
studied two-flavour ensembles \cite{Erben:2017hvr}. The implications
for $\ahvp$ will be discussed in a forthcoming publication
\cite{Erben:prep}. 

In Fig. \ref{fig:chiralcont} we show examples of our combined chiral
and continuum extrapolations for the charm and light quark
contributions. We perform a simultaneous extrapolation of the results
obtained from the local-conserved and local-local correlator to a
common value at the physical point, based on the fit function
$f(m_\pi,\beta)=\alpha_1+\alpha_2 m_\pi^2 +\alpha_3 m_\pi^2\ln m_\pi^2
+\alpha_4 a^2$. One finds that the availability of two different
discretisations of the correlator leads to a much more reliable
extrapolation, in particular in the case of the charm quark
contribution for which discretisation effects are quite large. Owing
to the fact that the hopping parameter corresponding to the bare charm
quark mass has not been determined yet on ensemble E250, there is
currently no direct result at the physical pion mass. Restricting the
analysis to the connected contribution only, we obtain the following
estimates for the light, strange and charm quark contributions at the
physical point:
\be\label{eq:ahvp}
  (\ahvp)^{ud}=643(21)\cdot10^{-10},\quad
  (\ahvp)^s=52.6(2.5)\cdot10^{-10},\quad
  (\ahvp)^c=14.95(47)\cdot10^{-10}. 
\ee
The quoted statistical errors are dominated by the uncertainty in the
lattice scale.

\begin{figure}[t]
\vspace{-0.4cm}
\centering
\leavevmode
\includegraphics[width=0.39\textwidth,clip]{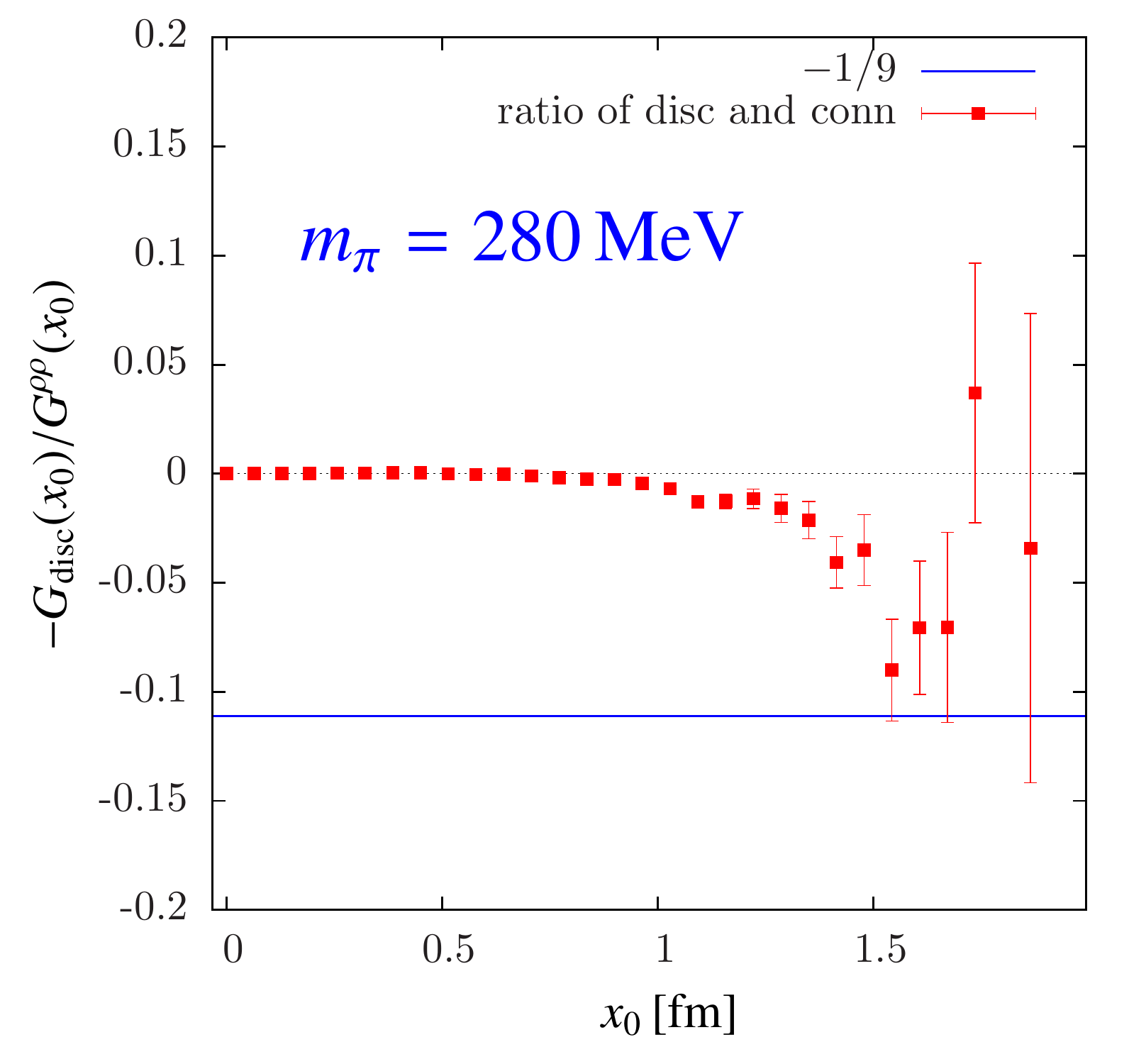}
\hfill
\includegraphics[width=0.54\textwidth,clip]{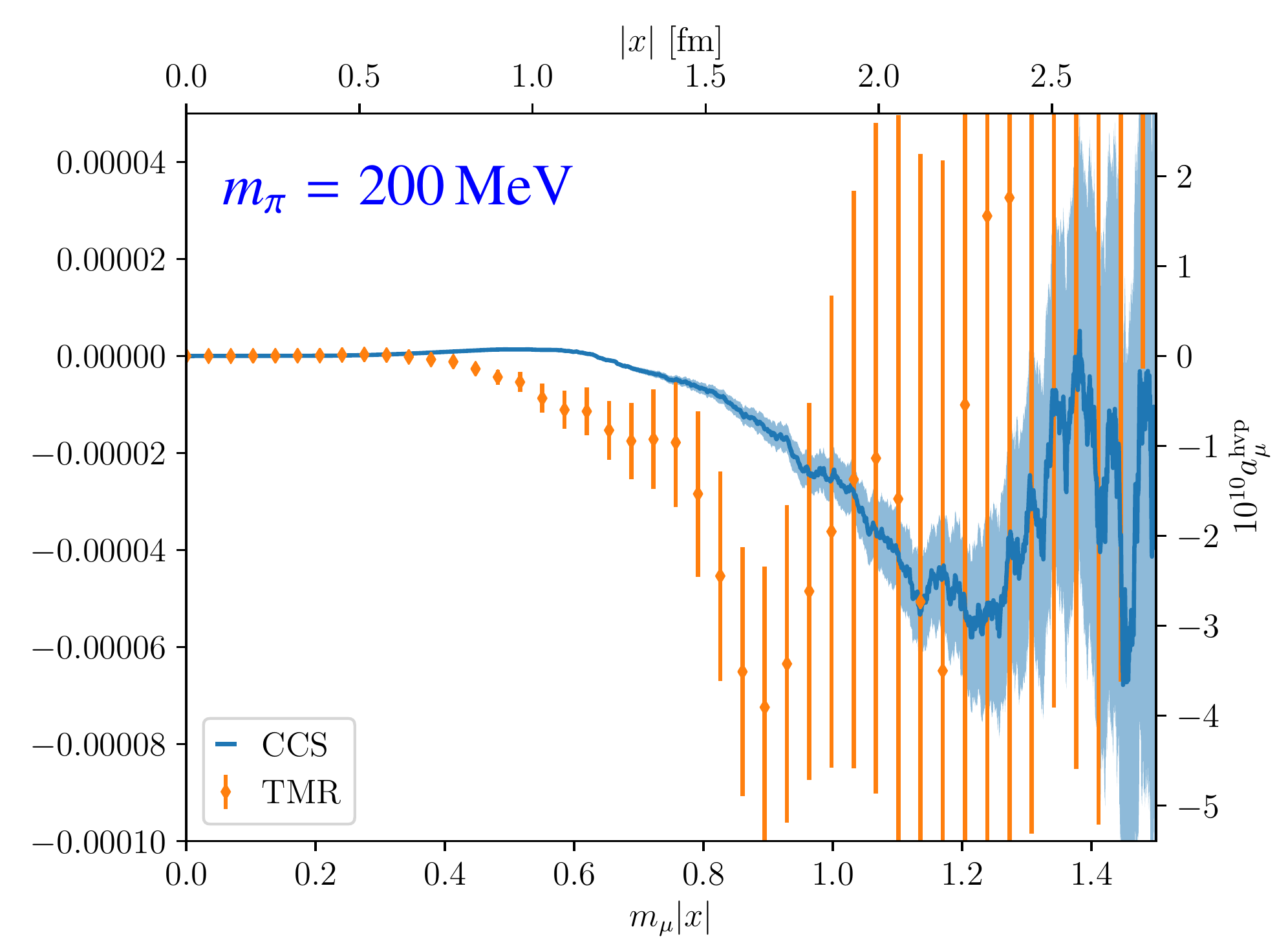}
\vspace{-0.3cm}
\caption{Left: The ratio of the disconnected over the connected light
  quark contribution (red circles) compared to the asymptotic value of
  $-1/9$ (blue horizontal line) for ensemble N200. Right: Comparison
  of the accumulated disconnected contribution determined using the
  TMR (orange circles) and the CCS method (blue band) for ensemble
  D200.\label{fig:disc}}
\end{figure}

We have also computed the contributions from quark-disconnected
diagrams, employing the technique of hierarchical probing with
Hadamard vectors \cite{Stathopoulos:2013aci}. As in our previous
calculation we took advantage of the cancellation of stochastic noise
between light and strange quark loops \cite{Francis:2014hoa}. In the
left panel of Fig. \ref{fig:disc} we plot the ratio
$-G_{\rm{disc}}(x_0)/G^{\rho\rho}(x_0)$ of the disconnected and the
light quark connected isovector contribution. As shown in
\cite{DellaMorte:2017dyu}, this ratio approaches the value $-1/9$ as
$x_0\to\infty$, and indeed we see the onset of the expected asymptotic
behaviour in the data. In the right panel we display a comparison of
the accumulated disconnected contribution to $\ahvp$
computed using our standard method to the results obtained using the
covariant coordinate space (CCS) method of Ref. \cite{Meyer:2017hjv},
which was also applied in a calculation of the hadronic corrections to
the weak mixing angle \cite{Ce:2018ziv}.\footnote{Data computed using
  the CCS method are plotted as a function of the upper integration
  bound $|x|^{\rm max}$. For the TMR method the values along the
  abscissa correspond to the integration interval over Euclidean
  time.} While both techniques give consistent results, the CCS method
is statistically more accurate.

\section{Conclusions}

Adding the contributions from light, strange and charm quarks in
Eq.\,(\ref{eq:ahvp}) we obtain our preliminary result of
$\ahvp=711\cdot 10^{-10}$, with an error of about 3\% which is
dominated by scale setting. Our calculation will be improved further
by increasing statistics, the inclusion of the contribution from
quark-disconnected diagrams and the effects from strong and
electromagnetic isospin breaking. Details of our calculational
framework can be found in Refs. \cite{Risch:2017xxe,Risch:2018ozp}.

\smallskip
\par\noindent{\bf Acknowledgments:} Our calculations were performed on
the HPC Clusters Wilson, Clover and MOGON-II at the University of
Mainz, on the JUQUEEN computer at NIC, J\"ulich (Project HMZ21) and on
Hazel Hen at HLRS, Stuttgart (Project GCS-HQCD, Acid 44131). We thank
Marco C\`e for providing the plot of the right-hand panel in
Fig.\,\ref{fig:disc}. KO is supported by DFG grant HI\,2014/1-1. We
thank our colleagues within the CLS initiative for sharing ensembles.

\providecommand{\href}[2]{#2}\begingroup\raggedright\endgroup

\end{document}